\documentclass[12pt,preprint]{aastex}

\sfcode`A=1000 \sfcode`B=1000 \sfcode`C=1000 \sfcode`D=1000
\sfcode`E=1000 \sfcode`F=1000 \sfcode`G=1000 \sfcode`H=1000
\sfcode`I=1000 \sfcode`J=1000 \sfcode`K=1000 \sfcode`L=1000
\sfcode`M=1000 \sfcode`N=1000 \sfcode`O=1000 \sfcode`P=1000
\sfcode`Q=1000 \sfcode`R=1000 \sfcode`S=1000 \sfcode`T=1000
\sfcode`U=1000 \sfcode`V=1000 \sfcode`W=1000 \sfcode`X=1000
\sfcode`Y=1000 \sfcode`Z=1000

\newcommand{\js}{}
\newcommand{\etal}{{\js et~al.\/}}

\newcommand{\Lsun}{\hbox{L$_{\odot}$}}
\newcommand{\Msun}{\hbox{M$_{\odot}$}}
\newcommand{\no}{\nodata} 
\newcommand{\mmjy}{\hbox{$\mu$Jy}}
\newcommand{\apphot}{{\sc apphot}}
\newcommand{\Subaru}{{Subaru}}
\newcommand{\Spitzer}{{\it Spitzer\/}}
\newcommand{\hst}{{\it HST}}

\slugcomment{Accepted by ApJ}
\shorttitle{AEGIS20 Identifications}
\shortauthors{Willner et al.}

\begin{document}


\title{A Fully-Identified  Sample of AEGIS20 Microjansky Radio Sources}


\author{
S.~P.~Willner,\altaffilmark{1}
M.~L.~N.~Ashby,\altaffilmark{1}
P.~Barmby,\altaffilmark{2}
S.~C.~Chapman,\altaffilmark{3}
A.~L.~Coil,\altaffilmark{4}
M.~C.~Cooper,\altaffilmark{5}
J.-S.~Huang,\altaffilmark{1}
R.~Ivison,\altaffilmark{6,7}
and
D.~C.~Koo,\altaffilmark{8}
}
\altaffiltext{1}{Harvard-Smithsonian Center for Astrophysics, 60 Garden Street,
Cambridge, MA 02138}
\altaffiltext{2}{University of Western Ontario, Dept.\ of Physics \&
  Astronomy, London, ON, Canada N6A 3K7}
\altaffiltext{3}{Institute of Astronomy, University of Cambridge,
Madingley Road, Cambridge, UK CB3 0HA}
\altaffiltext{4}{Department of Physics, University of California at
  San Diego, 9500 Gilman Dr., La Jolla, CA 92093} 
\altaffiltext{5}{University of California, Irvine, Dept.\ of Physics
  \& Astronomy, 4129 Reines Hall, Irvine, CA 92697}
\altaffiltext{6}{Astronomy Technology Centre, Royal Observatory,
Blackford Hill, Edinburgh EH9 3HJ UK}
\altaffiltext{7}{Institute for Astronomy, University of Edinburgh,
Blackford Hill, Edinburgh EH9 3HJ UK}
\altaffiltext{8}{UCO/Lick Observatory, Dept. of Astronomy \&
  Astrophysics, Univ. of California, Santa Cruz, CA 95064}


\begin{abstract}
  Infrared 3.6 to 8~\micron\ images of the Extended Groth Strip yield
  plausible counterpart identifications for all but one of 510 radio
  sources in the AEGIS20 $S(1.4~{\rm GHz})>50$~\mmjy\ sample.  This
  is the first such deep sample that has been effectively 100\%
  identified. Achieving the same identification rate at $R$-band
  would require observations reaching $R_{AB}>27$. Spectroscopic
  redshifts are available for 46\% of the sample and photometric
  redshifts for an additional 47\%.  Almost all of the sources with
  3.6~\micron\ AB magnitudes brighter than 19 have spectroscopic
  redshifts $z<1.1$, while fainter objects predominantly have
  photometric redshifts with $1\la z\la3$.  Unlike more powerful
  radio sources that are hosted by galaxies having large stellar
  masses within a relatively narrow range, the AEGIS20 counterparts
  have stellar masses spanning more than a factor of 10 at $z\sim1$.  The
  sources are roughly 10--15\% starbursts at $z\la0.5$ and 20--25\%
  AGNs mostly at $z>1$ with the remainder of uncertain nature.
\end{abstract}





\section{Introduction}

Radio observations are an excellent way to identify star-forming
galaxies and active galactic nuclei (AGNs).  Radio surveys are not 
subject to selection effects of obscuration or 
spectral line contamination, which affect visible-light surveys.
Even very distant objects can have 
large radio flux densities.  However, radio surveys alone are not
sufficient to understand the populations, and followup observations
are often more difficult than the initial radio survey.  The problem
is that redshift $z>1$ sources are  faint in visible light and
require very deep followup studies in order to achieve
identifications.

Despite the difficulties of counterpart identification, there are now
several radio samples with identification rates of
$\ga$90\%. \citet{Waddington2000} found optical counterparts for 96\%
of the sources in a $S(1.4~{\rm GHz})>1$~mJy sample with images
reaching $R=26$. \citet{Ciliegi2003} found counterparts for 92\% of
$S(6~{\rm cm})>50$~\mmjy\ sources with $I_{\rm AB}=25$
images. \citet{Afonso2006} were able to identify only 89\% of an
$S(1.4~{\rm GHz})>61$~\mmjy\ radio sample even with \hst/ACS
observations reaching magnitude $z_{850}=28$. \citet{Simpson2006}
identified $>$90\% of a 100~\mmjy\ radio sample using $BRi'z'$ images
reaching AB magnitude 27. \citet{Mainieri2008}, using data over a
wide wavelength range including the infrared, chose counterparts for
95\% of a 42--125~\mmjy\ radio sample but with an estimated 3\% rate
of spurious identifications.
In other recent work, \citet{Bardelli2010} reported an 82\%
identification rate in a $\sim$50~\mmjy\ radio sample in the COSMOS
field, and \citet{Afonso2011} achieved 83\% identification of a
sample of ultra-steep-spectrum radio sources with $S(610~{\rm
  MHz})>100$~\mmjy\ in the Lockman Hole. \citet{Huynh2008} identified
79\% of a much fainter ($S(1.4~{\rm GHz})>10$~\mmjy) radio sample
using deep HST images ($I_{\rm AB}<26$) albeit with a relatively
large matching radius (up to 1\farcs96).

Even when counterparts are detected, observed visible light corresponds
to rest-frame ultraviolet for high redshift galaxies.  While this can
give a measure of star formation {\em rate}, it gives little
indication of stellar {\em mass} and thus little indication of the
type of galaxy hosting the radio source.
Radio-quiet AGNs and star-forming galaxies both contribute to the
faint radio population, but lack of complete identifications  and 
limited wavelength coverage make
the proportions uncertain \citep[e.g.,][]{Huynh2008}.

The All-wavelength Extended Groth strip International Survey (AEGIS)
\citep{Davis2007} offers an unprecedented combination of deep,
multiwavelength data over a large area, the Extended Groth Strip
(EGS).  The data include a radio survey at 20~cm, AEGIS20
\citep{Ivison2006}, which reaches a sensitivity limit of
50~\mmjy~beam$^{-1}$. \citet{Willner2006} showed that infrared
observations of radio sources can produce very high identification
rates for radio source counterparts albeit at much higher flux
densities (55~mJy~beam$^{-1}$) than AEGIS20 and at 6~cm rather than
20~cm. \citet{Park2008} claimed to find all the AEGIS20 radio sources
that are also 24~\micron\ sources on the IRAC images but used a
2\farcs5 matching radius. IRAC data at 20~minute depth contributed to
the \citet{Afonso2011} identifications in the Lockman Hole.  Infrared
data should give high identification rates because the SED of typical
stellar populations peaks near 1.6~\micron.  For a distant source,
this peak will be redshifted to longer wavelength, and the IRAC 3.6
and 4.5~\micron\ flux densities will not decrease as rapidly as might
be expected.  Passive
evolution also increases the observed flux 
densities in these band: a stellar population of a given mass was
brighter in the past when it was younger.  (See Fig.~1 of
\citealt{Eisenhardt2008}.)  Thus counterparts should be visible in
IRAC observations unless they are either extremely distant, have very
low mass, or are heavily obscured by dust, neither of the latter two
being likely for a powerful radio galaxy.

This paper reports  the matching of AEGIS20 radio sources
primarily to IRAC 3.6 to 8.0~\micron\ data.
\citet{Barmby2008} have provided images and catalogs of IRAC data in
the EGS.  The typical exposure time is
2.5~hours (9~ks), and the 80\% completeness limits for the catalog are
$\sim$5~\mmjy\ at 3.6 and 4.5~\micron\ and $\sim$10~\mmjy\ at 5.8
and 8.0~\micron.  However, fainter objects with known positions can
be identified on the images. Additional IRAC data for the EGS exist
(Ashby et al., in preparation, 2012) but were not used for this
work. In practice, the \citeauthor{Barmby2008} data suffice to
identify counterparts for all or nearly all of the radio sources.
The radio sample is defined and source matching is described in
Section~\ref{s:id}, counterpart properties including photometry and
redshifts are given in Section~\ref{s:prop}, and results are
discussed in Section~\ref{s:disc} and summarized in Section~\ref{s:conc}.
Throughout this paper, magnitudes are in the AB system, and the 
notation $[w]$ means the AB magnitude at wavelength $w$ in
\micron.\footnote{Some other papers use the $[w]$ notation to mean
  Vega magnitudes, but 
  here it means AB.}
Source distances are based on standard $\Lambda$CDM cosmology with
$H_0=71$~km~s$^{-1}$~Mpc and $\Omega_M = 0.27$.  Practical calculation
of luminosity distances was based on the  program {\sc angsiz}
\citep{Kayser1997}.

\section{Sample Definition and Identifications}
\label{s:id}

\subsection{Selecting Counterparts with IRAC data}
\label{ss:id}

The initial radio catalog \citep{Ivison2006} contains 1122
sources\footnote{One source in the original catalog is a duplicate.}
with $S(1.4~{\rm GHz})>50$~\mmjy.  Of these, 511 are in the area of
IRAC coverage, defined by at least five independent images in each of
the four IRAC channels.  (This amounts to one-tenth the nominal IRAC
exposure time, but most sources were observed with the nominal
exposure time or close to it.)  These 511 radio positions, within an
area of $\sim$950~arcmin$^2$,  are the
ones we have attempted to match to IRAC sources. 

\begin{figure}
\epsscale{0.5}
\plotone{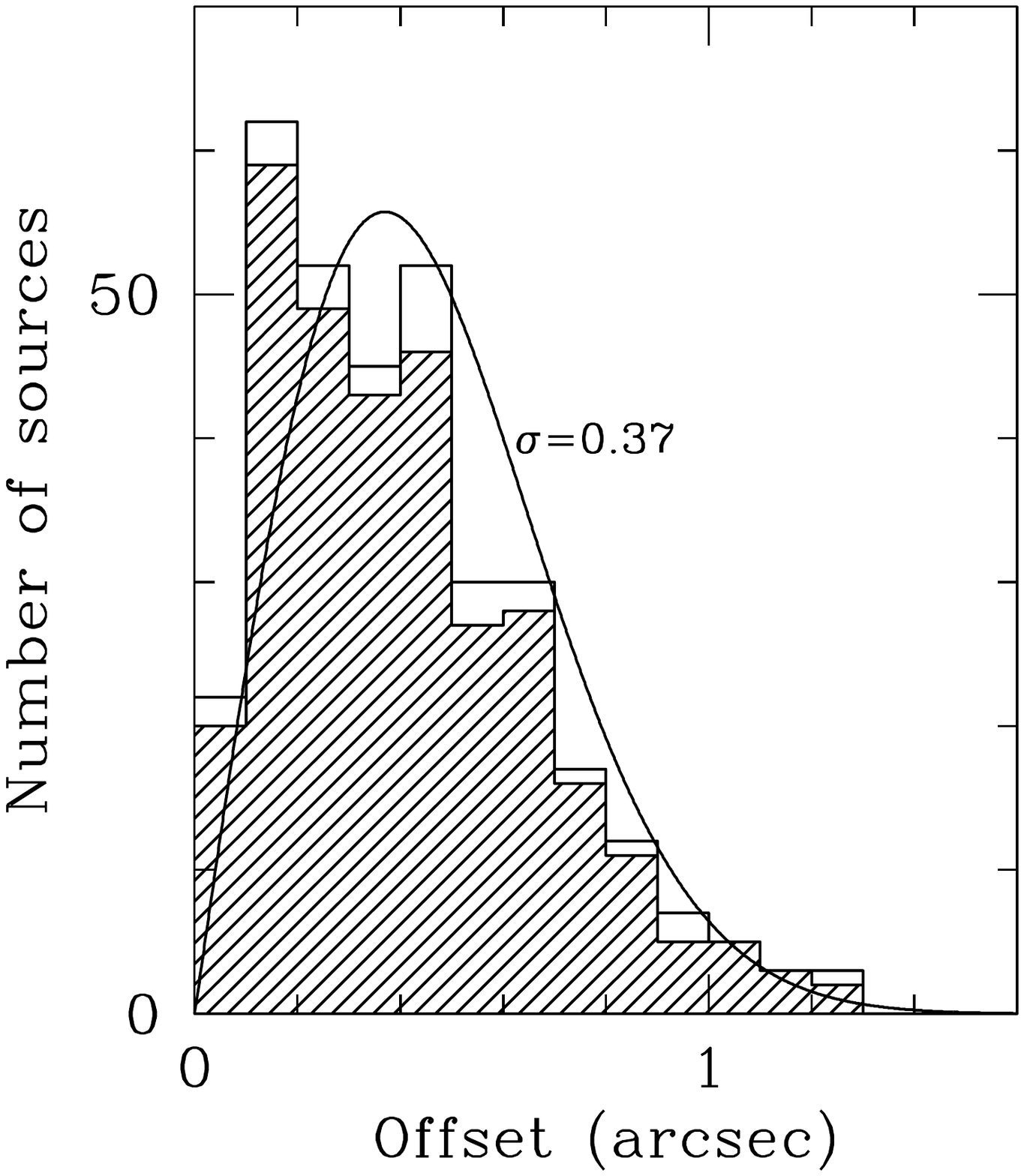}
\caption{Histogram of radial offsets between radio and IRAC positions
for 340 radio point sources.  The shaded region shows offsets for 314
sources with $[3.6]<22$.  The curve shows the expected distribution
for Gaussian position errors of 0\farcs37 \citep{Barmby2008} in each coordinate.}
\label{f:hist}
\end{figure}

The simplest radio sources to match were the point sources, those
unresolved at the 3\farcs8~FWHM of the radio beam
\citep{Ivison2006}. There are 342 such sources in the input list. One
is 3\farcs0 from the bright ($K_S=8.4$) star 2MASS J14230588+5333504.
No identification is possible for this one, and it was dropped from
further consideration.  Of the remaining 341 sources, all but one
coincide with an IRAC catalog source consistent with Gaussian
position uncertainties having a standard deviation of 0\farcs37 in
each coordinate.  There is no systematic offset in right ascension,
and the offset in declination is 0\farcs08, the IRAC positions being
systematically north of the radio positions. Figure~\ref{f:hist}
shows a histogram of the radial offsets. The one badly discrepant
source is 434 (Table~\ref{t:rad}), which has an offset of
1\farcs6.\footnote{Other sources with large offsets are 033 and 487
  with offsets of 1\farcs3.  The latter source is near the edge of
  the IRAC map and has only 1/10 normal coverage; its IRAC position
  uncertainty is thus expected to be about three times larger than
  normal.  Indeed a systematic shift is seen between the 3.6 and
  4.5~\micron\ images for this source. Source 33 has full-depth IRAC
  coverage with no obvious position error, but its offset could be
  the 3-sigma outlier expected in this sample.  It is a 5.4$\sigma$
  radio detection, thus unlikely to be spurious.}  This is too large
to be considered a valid identification, and if the radio source is
real, it remains unidentified.  The radio flux density is at the
catalog limit of 50~\mmjy, just under a 5$\sigma$ detection. Thus
340 of the 341 point sources possible to match are indeed
matched by catalog IRAC sources.

Most of the resolved radio sources are equally simple to match --- the
sources are small, and there is an obvious IRAC counterpart --- but some of
the complex, resolved sources present more of a challenge.  For seven radio
sources, the IRAC data suggest two possible counterparts.  Radio
images for these seven are elongated and statistically consistent
with two point sources separated by a few arcseconds.  In four of
these cases, detection of the weaker radio source was less than
3$\sigma$ significance.  In these cases, the radio source was deemed
to be a single point source, and the nearest galaxy to it was adopted
as the counterpart.  The other galaxy might be a radio source as
well, but its flux density is below the sensitivity limit of the AEGIS20
survey.  In the other three cases, both radio components were
detected at $>$3$\sigma$, and these are treated as separate sources,
each radio component having an IRAC counterpart.  Radio to IRAC
offsets are consistent with the position uncertainties found for the
point sources.

\begin{figure}
\vspace*{-2in}
\includegraphics[width=\textwidth]{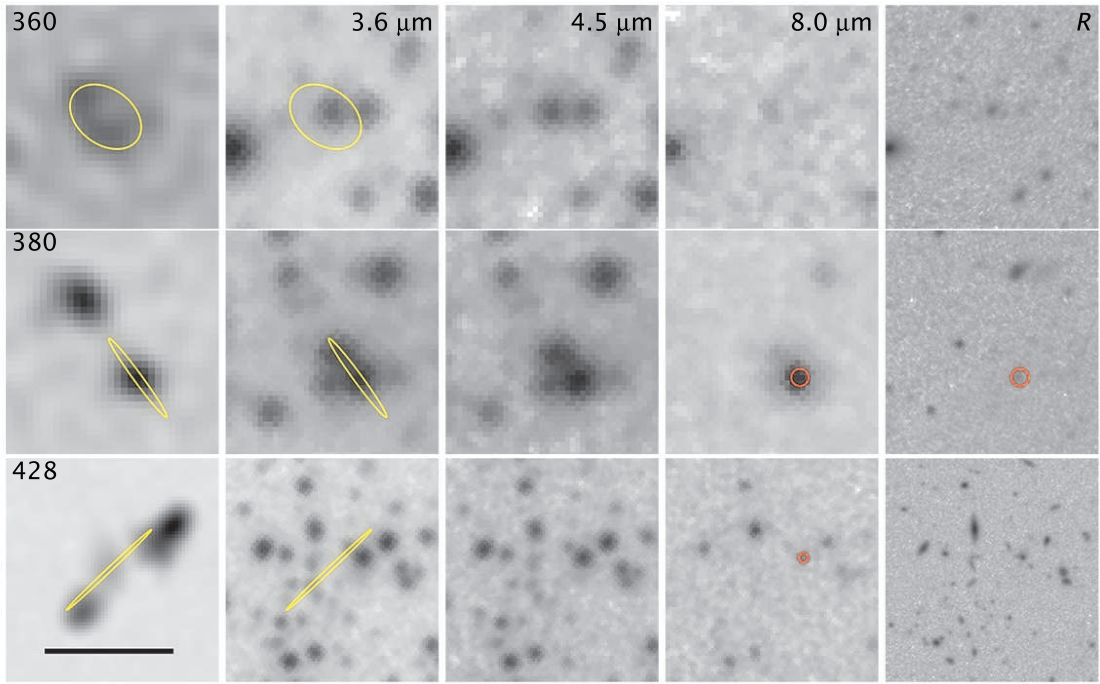}
\caption{Postage stamp images of complex sources.  From left to
  right, each row shows the VLA 20~cm radio, the IRAC 3.6~\micron,
  4.5~\micron, and 8.0~\micron\ images and the Subaru $R$ image of a
  source.  The source numbers from Table~\ref{t:rad} are indicated in
  the left panels.  Scale bars indicate 30\arcsec; each stamp image
  is 100\arcsec, 50\arcsec, or if no scale bar is shown,
  25\arcsec\ across.  North is up and east to the left in all
  images. Yellow contours show the Gaussian fit \citep{Ivison2006} to
  the deconvolved radio source.  The fits are generally smaller than
  the radio images themselves because of the radio beam size. Red
  circles are 2\arcsec\ in diameter and show the position of the IRAC
  counterpart where it might not be obvious.}
\label{f:stamps}
\end{figure}

Finally, there are 15 radio sources with complex morphologies, shown
in Figure~\ref{f:stamps}.  All but one have a plausible IRAC
counterpart, though in some cases there are multiple candidates, and
there is no way to be certain we have chosen the correct one.  For
source 002, there is a bright IRAC source within 5\farcs5 of the
radio peak, but a more plausible explanation is that source 002 is
the tip of a jet emerging from source 003.  There is a counterjet in
the opposite direction, though it is not bright enough in the radio
image to have been detected by the automated search as a separate
source.  The extended ($\sim$5\arcsec) radio source 236 is 4\farcs3
from our suggested IRAC conterpart (an $R=24.4$ galaxy at $z_{\rm
  phot}=0.69$) corresponding to a projected separation of about
30~kpc.  Alternative IRAC identifications  are a fainter galaxy 
(shown at the northern tip of the radio ellipse in
Fig.~\ref{f:stamps}) 3\farcs4 from the radio
centroid\footnote{$R=24.5$, $z_{\rm phot} = 0.6$ according to the
  Rainbow database.} or that 236 is a radio lobe of source 233 to the
southeast.  Source 233 has a radio jet and lobe extending to its
southwest, but the radio image shows no evidence of any jet
connecting 233 to 236.  Source 428 has multiple radio lobes, and the
counterpart could be any of many sources
visible in the IRAC image.  We suggest the most likely counterpart is
a $z=0.8260$, $R=22.7$ galaxy bright at 8~\micron\ and located near the inner
edge of the northwest radio lobe.  At the suggested redshift, the projected
42\arcsec\  maximum extent of the radio lobes corresponds to a
length of 320~kpc.

The overall identification rate is remarkably high.  At most three of
510 radio sources lack IRAC counterparts, and there is a plausible
case that the identification rate is 100\%.  This would require
source 2 to be a jet from 3 (as is likely), source 236 to be related
to one of two candidate counterparts or be a jet from 233, and
AEGIS20 radio source 434 to 
be spurious.  At the near-5$\sigma$ detection level,
even a single 
spurious source is statistically unlikely, but the absence of an IRAC
counterpart is hard to explain.  At $z=1$, the IRAC
3.6~\micron\ detection limit 0.7~\mmjy\footnote{This limit is for the
  nominal survey exposure time of 9~ks.  The actual exposure time at
  the position of 434 was 3.4~ks at 3.6~\micron\ and 6.6~ks at
  4.5~\micron, so the actual limits for this source are correspondingly worse.}
corresponds to a stellar
luminosity of order $10^9$~\Lsun, and such low luminosity galaxies
are unlikely to harbor powerful radio sources.  Even at $z=5$, a
galaxy with $L=10^{11}$~\Lsun\ should have been detected at
4.5~\micron.  Nevertheless, \citet{Huynh2010a} reported two
radio sources in the Extended {\it Chandra} Deep Field South without
apparent IRAC counterparts.  If such a 
population exists, source  434 (and less likely source 236) could be
a member.  The source
density suggested by \citeauthor{Huynh2010a} predicts $>$4 such
sources in the 
EGS field, where we find at most two and maybe none. 
This implies radio sources without IRAC counterparts are rare if they
exist at all.

\subsection{Selecting Counterparts with $R$-band data}

\begin{figure}
\vspace*{-2in}
\includegraphics[width=\textwidth]{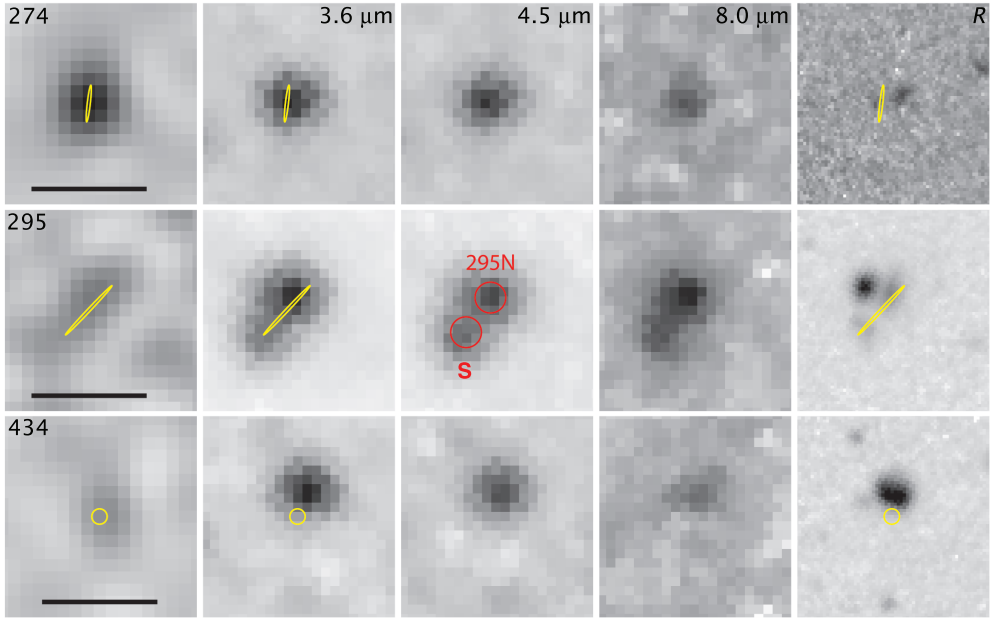}\protect
\caption{Postage stamp images of counterparts that might be
  misidentified if only $R$ images were available and one source
  possibly lacking an IRAC counterpart.  Each row shows one
  source with number (Table~\ref{t:rad}) indicated in the left
  panel.  Panels from left to right show the
  20-cm radio image,  IRAC 3.6~\micron, 4.5~\micron, and
  8.0~\micron\ images, and the Subaru $R$ image.  Scale bars indicate
  7\farcs5; each stamp image is 12\farcs5 across.   North is up and east to the
  left in all images.  Yellow contours show the Gaussian fit
  \citep{Ivison2006} to the deconvolved radio source or a 1\arcsec\ diameter
  circle if the radio source was unresolved. Red circles are 1\arcsec\
  in diameter and show the position(s) of the IRAC counterpart(s).}
\label{f:stamps2}
\end{figure}

Deep \Subaru\ $R$-band data exist for almost the entire EGS\footnote{
  Five counterparts in the northern part of the EGS are outside the
  coverage of the $R$ images.} with 5$\sigma$ depths of 26.5 AB
magnitudes over most of the strip and 26.1 in the southwest portion.
These images were searched with SExtractor \citep{Bertin1996}, and
405 counterparts were found within 1\arcsec\ of the IRAC position.
The number would be 372 if a more stringent (and more reasonable)
coincidence criterion $<$0\farcs7 were used.\footnote{The search
  was based on IRAC positions because the purpose was to determine
  whether an apparent $R$ counterpart was the same object as the IRAC
  source. A purely visible search for radio source counterparts
  would, of course, start from VLA positions rather than IRAC
  positions and would probably find slightly fewer counterparts than
  the numbers stated.} This detection rate of ${<}$80\% (or $<$73\%)
is typical of visible-light counterpart searches (\S1).  Moreover,
for 10 radio sources, the automated $R$ search found a
neighbor object brighter in $R$ instead of the object detected by
IRAC.  We regard the latter as more plausible counterparts because
most radio sources originate in massive galaxies, which may be red
but will seldom have very blue colors, as would be required if any of
these 10 $R$ sources is the correct counterpart. Therefore about
2--3\% of the proposed counterparts would be incorrect in the
1\arcsec\ search, though all of these would be eliminated with the
0\farcs7 search.  \citet{Mainieri2008} also gave examples of
counterparts found in infrared $K_s$ band or IRAC data but invisible
at shorter wavelengths.  Figure~\ref{f:stamps2} shows images of some
sources that an automated $R$-band search might misidentify.


\section{Counterpart properties}
\label{s:prop}

\subsection{Source photometry}

No single method of photometry is adequate for all objects.  With
deep images, many radio counterparts have nearby  sources that
can contribute to the photometry in large apertures.  Other
counterparts are significantly extended, and only large apertures can
capture all the flux.  There are technical problems in some cases
because an automated aperture selection can combine unrelated objects
or separate a single object into multiple ones, though the latter
doesn't seem to have occurred in the present sample.  In practice, we
have measured IRAC and $R$-band fluxes through four apertures for each object: three
circular apertures with radii 1\farcs53, 2\farcs14, and
3\farcs06\footnote{These
aperture sizes were those used in the IRAC catalog
\citep{Barmby2008} and correspond to 2.5, 3.5, and 5.0 IRAC mosaic
pixels.  The IRAC mosaics have pixels half the size of the actual
IRAC pixels.}  and
the Kron aperture chosen by SExtractor ({\sc mag\_auto}).   

The default photometry aperture was the smallest circular one because of its
rejection of neighboring sources and sky fluctuations, but all
galaxies with semimajor axis size $>$1\farcs22 (measured by
SExtractor in $R$) were examined individually on the images, as were
objects with $m(1\farcs53)-m(2\farcs14)>0.25$ or
$|m(3\farcs06)-m_{\rm auto}|>0.25$.  In each case, the most
reasonable magnitude to include the whole galaxy but exclude neighbor
objects was chosen. This process was necessarily somewhat subjective,
but no alternative reasonable choice would change the magnitude by
more than about 0.2~mag.  In all, 166 counterparts were examined
individually to choose apertures.

SExtractor \citep{Bertin1996} was used to measure
both circular and {\sc auto} magnitudes for the sources it detected;
the SExtractor IRAC magnitudes were published by
\citet{Barmby2008}.\footnote{Source 036 is  not in the published catalog, but
  its {\sc auto} magnitudes were measured on the images in the same
  way as for the catalog.  Extended source
  corrections of 0.062, 0.040, 0.137, and 0.212, based on
  the latest coefficients from the \Spitzer\ Science center, are included
  in the Table~\ref{t:iracmags} magnitudes.  The channel~2 correction differs by
  0.016~mag from the one that would have been used in 2008, but 036 is the only radio
  source counterpart that needs an extended source correction.}
For sources not detected by the automated
SExtractor search, positions were measured from the images and
aperture photometry done at those positions with the IRAF
task {\tt apphot}.  The $R$ magnitudes were calibrated from SDSS
stars (avoiding those saturated in the Subaru data), for which
$R_{\rm Subaru} = r_{\rm Sloan} + 0.17(r-i)_{\rm Sloan}$ with 
rms scatter $\sim$0.05~mag.\footnote{The color term is poorly
  determined because the $r-i$ color covers only the range
  0--2~mag.}  The $R$ magnitudes are  
given in Table~\ref{t:rmags}, and IRAC magnitudes are given in
Table~\ref{t:iracmags}, which uses SExtractor magnitudes when they
exist and {\apphot} magnitudes otherwise. When {\sc auto} magnitudes
are used, the Kron apertures were derived separately in the IRAC
bands and in $R$ and may differ, but in both cases they are intended to
measure the total magnitude of the object.  The
uncertainties listed in Tables~\ref{t:rmags} and~\ref{t:iracmags} are
statistical only as derived from the local background fluctuations.
Upper limits are 5$\sigma$.

The $R$ magnitudes were checked in two ways.  The \Subaru\ data came
from three adjacent images, which overlap enough to include duplicate
images of 26 sources not strongly affected by artifacts or image
edges.  The rms absolute difference of observations of the same
source in the 1\farcs53 aperture is 0.04~mag, and the maximum is
0.10~mag. However, a difference of 0.86~mag was found for one source
near an image edge (one image being obviously bad at that location),
and a very few 
sources  near image edges or with other problems might have
bad magnitudes that have not been noticed. The rms difference between
SExtractor and {\sc 
  apphot} 1\farcs53 aperture magnitudes is 0.09~mag, presumably due
to differences in aperture centering and background
calculations. This is a reasonable lower limit on systematic
uncertainty for the fainter sources. However, it seems prudent to
recognize that in some cases the source measured in $R$ may be a
different object than the one measured by IRAC, and a few large
errors are possible.

\begin{figure}
\epsscale{0.5}
\plotone{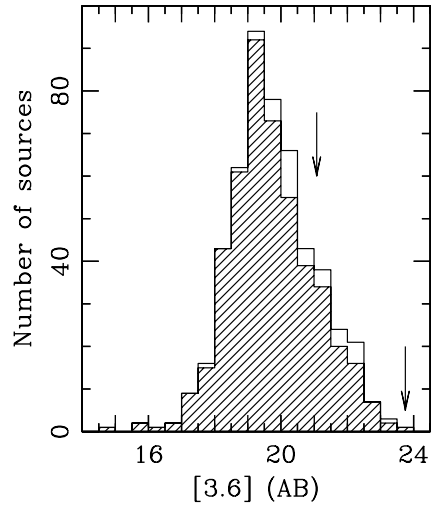}
\caption{Histogram of 3.6~\micron\ magnitudes of radio source counterparts.
  The hatched histogram shows magnitudes for sources with
  redshifts.  Arrows show the 90\% and 50\% completeness limits of
  the \citet{Barmby2008} catalog, but magnitudes can be measured on
  the images to fainter limits than these. Bins are half a magnitude
  in width because the distribution of 3.6~\micron\ magnitudes is
  narrower than the distribution of $R$ magnitudes (Fig.~\ref{f:rhist}).}
\label{f:mhist}
\end{figure}

\begin{figure}
\epsscale{0.5}
\plotone{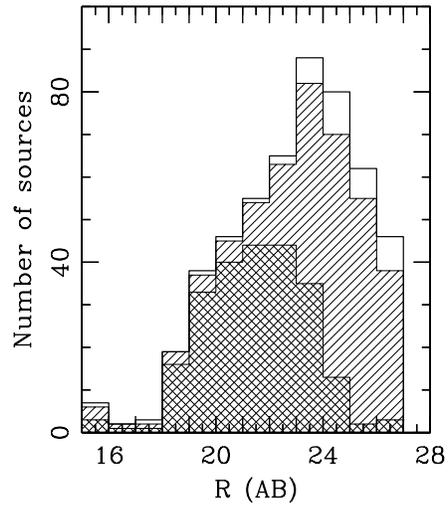}
\caption{Histogram of $R$ magnitudes of radio source counterparts.
  The hatched histogram shows magnitudes for sources with redshifts
  (including photometric ones),
  and the cross-hatched histogram for sources with spectroscopic
  redshifts.   When no  source was detected at $R$,
  its magnitude was placed in the bin corresponding to the
  upper limit. Some sources are likely much fainter than $R=27$.}
\label{f:rhist}
\end{figure}

The new {\tt apphot} IRAC aperture measurements agree with the
published ones except when there are nearby, confusing sources.  In
some of those cases, SExtractor can separate the sources, whereas
{\tt apphot} simply adds all the counts in the defined
aperture. Otherwise, testing showed that SExtractor and {\tt apphot}
magnitudes are consistent with each other.  \citet{Barmby2008}
discussed the uncertainties in the IRAC data, and the new
measurements should have equivalent uncertainties.
Figures~\ref{f:mhist} and~\ref{f:rhist} show the magnitude
distributions of the radio source counterparts in $[3.6]$ and $R$,
respectively.

\subsection{Source Redshifts}

Spectroscopic redshifts come from a variety of sources.  The largest
number (125) are from the DEEP2 redshift survey
\citep{Davis2007,Newman2012}. Additional redshift sources are listed in notes to
Table~\ref{t:rmags}.  All in all, 235 counterparts have spectroscopic
redshifts, and agreement is excellent  for the 44 sources with more
than one spectroscopic measurement.\footnote{$\Delta z \le 0.0015$ except
for one source with $\Delta z = 0.0054$.}

\citet{Barro2011a} have compiled a database of photometric and
spectroscopic data\footnote{ The Rainbow database can be found at
  https://rainbowx.fis.ucm.es/Rainbow\_navigator\_public/ .} in the EGS
and used it to derive photometric redshifts \citep{Barro2011b}.
Searching their database using the IRAC positions
(Table~\ref{t:iracpos}) of the radio source counterparts and
eliminating duplicates resulted in 440 matches within 0\farcs65.  We
examined  images and found
8 additional cases where the Rainbow object
appears  to be the one we identify as the radio
source counterpart;  position offsets for these were between
0\farcs65 and 0\farcs80.  No objects with offsets $>$0\farcs8
appeared to be valid matches. Of the 448 matched objects, 215 have spectroscopic
redshifts, and all but 21 agree with the \citet{Barro2011b}
photometric redshifts within $\Delta z \le 0.2$.  Another large
source of photometric redshifts for the EGS is the CFHT Legacy Survey
\citep[CFHTLS,][]{Coupon2009} with 183 altogether of which 105 are for
objects with spectroscopic redshifts.  Of these, 94 photometric
redshifts are within $\Delta z\le0.2$ of the spectroscopic ones,
again giving a success rate near 90\%.  The next largest source is
the NEWFIRM Medium Band Survey \citep[NMBS][]{Whitaker2011} with 76
redshifts. Of these, 5 of 42 differ from a spectroscopic measurement
by $\Delta z>0.2$.  Finally, 236 photometric redshifts are from
unpublished work \citep{Huang2012b}.  These use a neural network
technique that is inherently limited to $z\le1.1$ because of
insufficient galaxies at larger redshifts in the training
set. Nevertheless, of 71 with spectroscopic measurements, only 11
deviate by more than $\Delta z=0.2$.  The overall results suggest a
photometric redshift success rate near 90\%, but sources lacking
spectroscopic redshifts tend on average to be fainter, and the
success rate could be lower for them.  Table~\ref{t:rmags} gives the
best redshift we could find for each object.  We have preferred NMBS
if available because the use of medium band filters should give
better redshift performance.  The least-preferred survey was the one
from \citet{Huang2012b} because of its limitation to $z<1.1$, but order
of preference among the other surveys was not obvious.  Changing the
order would not change the results of this paper.  All in all,
Table~\ref{t:rmags} has 235 spectroscopic redshifts, 238 photometric
redshifts, and 38 objects with no redshift available.  Thus about
93\% of the radio source counterparts have a redshift of some kind,
but at least 24 of the photometric redshifts (5\% of the full sample)
likely differ from the true redshift by $\Delta z > 0.2$.

\section{Discussion}
\label{s:disc}

\subsection{Identification Rate}

The overall identification rate of $>$99\% is unprecedented for a
radio sample of this size and depth, and the sensitive infrared data
from IRAC were critical.  87 counterparts or 17\% of the sample are
fainter than the 90\% catalog completeness limit of 21.08~mag.  If
the IRAC images had been shallower, these sources would have been
unidentified or incorrectly matched to neighboring sources brighter
than the correct counterparts. Indeed, \citet{Afonso2011} achieved
only 83\% identification with their shallower IRAC data, though their
radio sample (consisting solely of ultra-steep-spectrum radio
sources) is not identical to ours and may preferentially include
higher-redshift objects.

Spurious counterpart identifications from chance coincidence of radio
sources with unrelated IRAC sources are unlikely. The IRAC source
density \citep[][Fig.~9]{Barmby2008} at $[3.6]<21$ is about
0.002~arcsec$^{-2}$.  Thus we would expect blind position matching of
the whole radio sample with radius 0\farcs7 to produce less than one
spurious match at this magnitude level.  In fact, 417 counterparts
have $[3.6]<21$ (Fig.~\ref{f:mhist}), and position offsets are mostly
$<$0\farcs7 (Fig.~\ref{f:hist}), consistent with expected position
uncertainties.  Of the fainter 93 sources, we expect $\sim$1 spurious
match within 0\farcs7 and $[3.6]<23$, but again the magnitudes are
nearly all brighter and the position offsets smaller than these
values.  The most suspicious source is 033 with $[3.6]=22.4$ and a
position offset 1\farcs3.  The proposed counterpart has a relatively
flat SED (in $F_\nu$, Table~\ref{t:iracmags}) from 3.6 to
8.0~\micron, unusual for a normal galaxy but not for a radio source
counterpart.  Thus we regard even a single spurious IRAC match for
the simple radio sources as unlikely. The areal density of $R<26$
sources is about 5 times higher than of $[3.6]<23$ sources, and a few
spurious matches at $R$ cannot be ruled out.  For the complex
sources, the issue is not so much chance coincidences as uncertainty
in the expected location of the counterpart relative to the radio
emission.  As discussed in Section~\ref{ss:id}, alternate counterpart
identifications are possible in some cases.

\subsection{Nature of Sources}

Radio emission can arise either from an AGN or from star formation
(e.g., \citealt{Rieke1980,Yun2001,Condon2002,Bell2003} and especially
\S2.1 of \citealt{Padovani2009} for comprehensive discussion).  There
are a variety of ways to separate the respective contributions
\citep[e.g.,][]{Padovani2009,Moric2010}, but in general 
detailed followup at wavelengths other than radio and mid-infrared
(MIR) is required.  

\begin{figure}
\epsscale{1.0}
\plotone{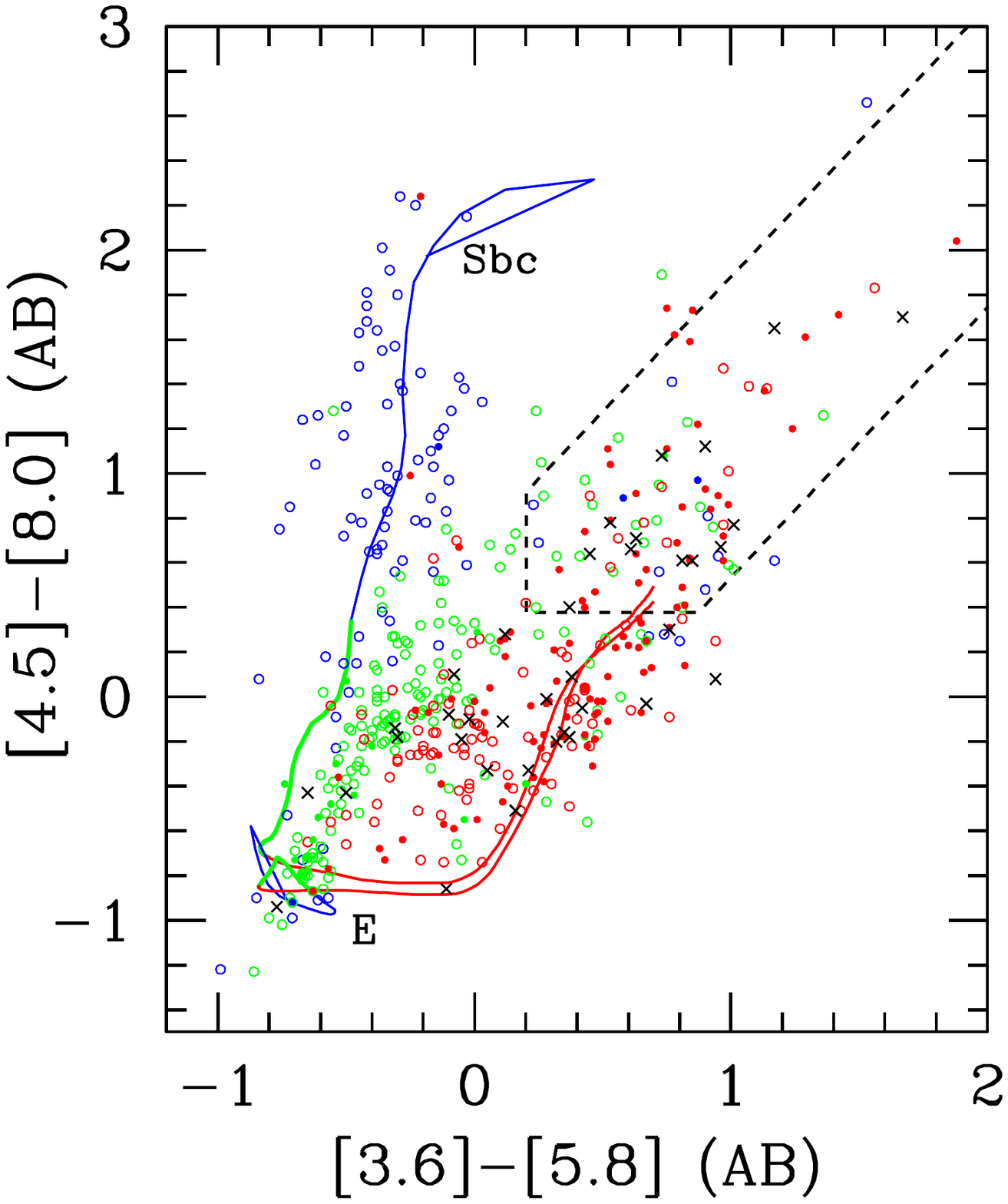}
\vspace*{-1.5in}
\caption{Color-color diagram for radio source counterparts showing the AGN selection
  region proposed by \citet{Donley2011}.  Circles denote objects with
  redshifts: blue for $z\le0.5$, green for $0.5<z\le1.1$, and red for
  $z>1.1$.  Circles are filled if the radio luminosity $L({\rm
    3~GHz}) >10^{23}$~W~Hz$^{-1}$ at $z\le0.5$ or
  $>10^{23}$~W~Hz$^{-1}$ at $z.0.5$. Crosses denote objects lacking
  redshifts.  Curves illustrate the colors of two template SEDs, one
  for an elliptical galaxy (E) and the other for an Sbc galaxy, as
  redshift increases. The curves are color coded by redshift in the
  same way as the points and labeled near their respective $z=0$ locations.}
\label{f:donley}
\end{figure}

\begin{figure}
\epsscale{1.0}
\plotone{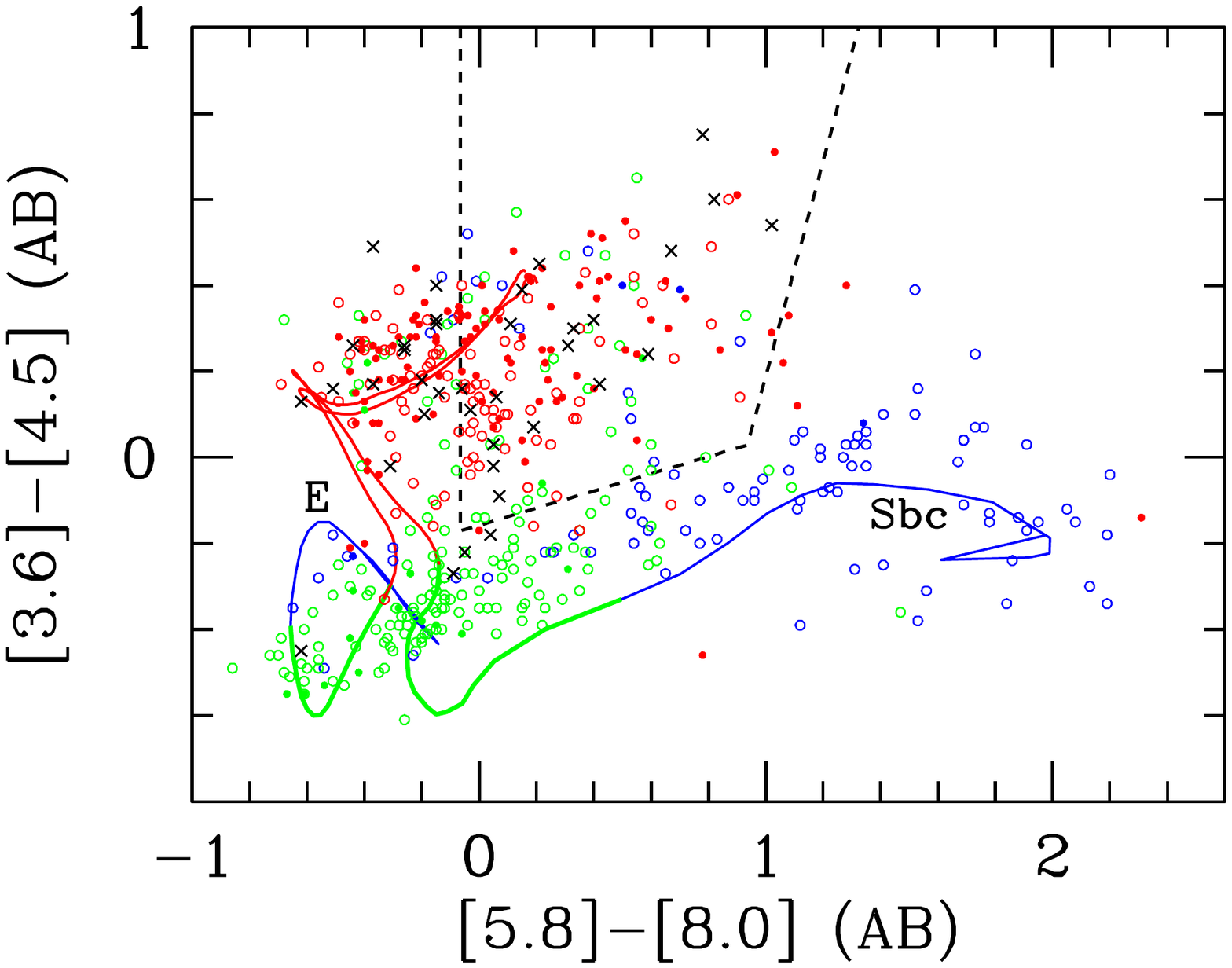}
\vspace*{-1.5in}
\caption{Color-color diagram for radio source counterparts showing
  the AGN selection wedge proposed by \citet{Stern2005}.  
  Circles denote objects with redshifts: blue for $z\le0.5$,  green for
  $0.5<z\le1.1$, and red for $z>1.1$.  Circles are filled if the radio luminosity $L({\rm
    3~GHz}) >10^{23}$~W~Hz$^{-1}$ at $z\le0.5$ or
  $>10^{23}$~W~Hz$^{-1}$ at $z.0.5$. Crosses denote objects lacking
  redshifts.  Curves
illustrate the colors of two template SEDs, one for an elliptical
galaxy (E) and the other for an Sbc galaxy, as redshift increases.
The curves are color coded by redshift in the same way as the points.}
\label{f:cc1}
\end{figure}

For some objects, the source of emission is obvious. The low-redshift
starburst population can be seen in Figure~\ref{f:donley}, where the
8~\micron\ PAH emission makes low-$z$ starbursts red in the
$[4.5]-[8.0]$ IRAC color (but does not affect $[3.6]-[5.8]$, which
remains blue) as long as $z\la0.5$. (See Fig.~8 of
\citealt{Huang2007}. Fig.~2 of \citealt{Padovani2011} is the
equivalent diagram for their sample.) Figure~\ref{f:cc1} shows a
different color-color plot.  In both plots, the low-redshift galaxies
are mostly within the starburst region, but a significant minority
are in the AGN region.  Depending on which color-color plot and the
exact criteria one chooses, there are about 50--60 $z\le0.5$ galaxies
in the low-$z$ starburst category.

Other cases where the emission source is obvious are those with
spectra that show strong PAH features. \citet{Huang2009} found 11
sources with $1.6\le z \le 3.0$ and strong PAH features; seven of
them are in the present radio sample.  A seventh radio source in the
\citeauthor{Huang2009} list, not in their original sample but
discovered serendipitously, shows no PAH features and is likely to be
an AGN at $z=2.12$ both from its spectrum and its MIR colors.

\begin{figure}
\epsscale{1.0}
\plotone{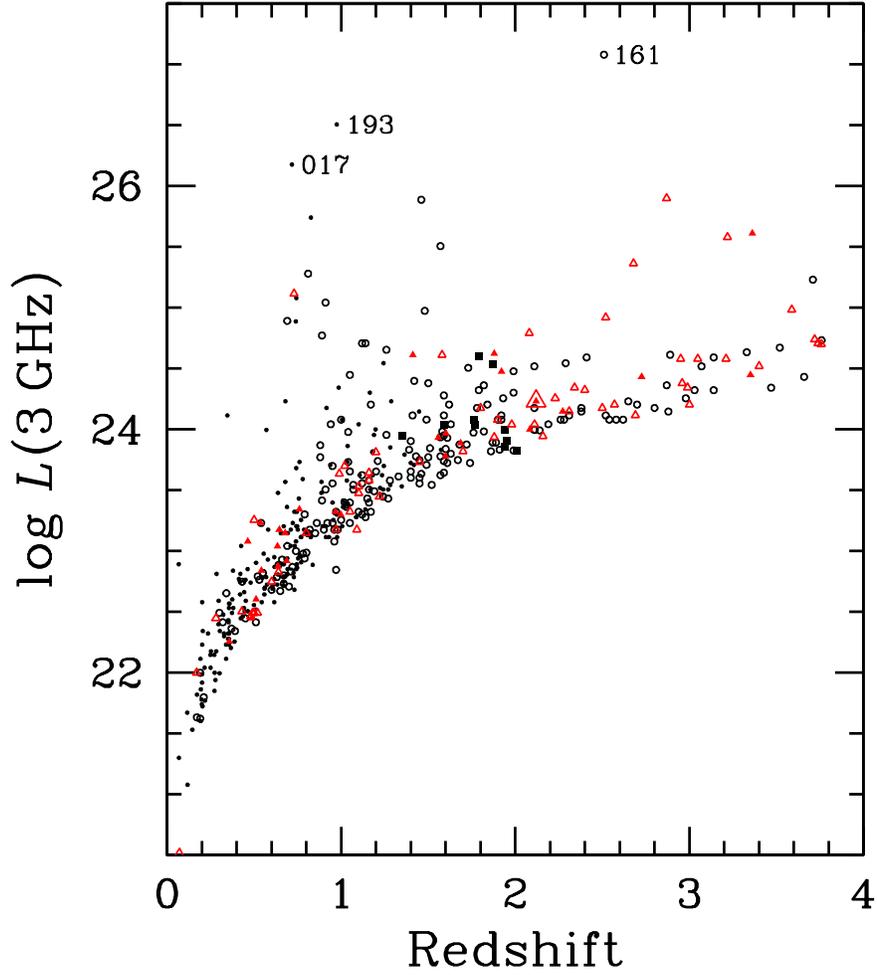}
\vspace*{-1.5in}
\caption{Radio power as a function of redshift for all sources that
  have redshifts.  Red triangles denote sources with MIR colors
  indicating AGN emission, and black circles denote other
  objects. Symbols are filled if the redshift is spectroscopic.
  Black squares denote ULIRG sources from
  \citet{Huang2009,Huang2012a}, and the large red triangle denotes the
  one AGN source from \citet{Huang2009}. Three extreme sources are
  labeled with source nicknames from Table~\ref{t:iracpos}.}
\label{f:zp}
\end{figure}

Radio power itself can be used as a criterion to separate star formation
from active nuclei in local galaxies \citep[e.g.,][]{Padovani2009}, but it
may not work well at large redshifts.  A radio luminosity (measured
for convenience at rest frequency 3~GHz) $L({\rm 3~GHz}) =
10^{23}$~W~Hz$^{-1}$ corresponds to SFR = 100~\Msun~yr$^{-1}$
\citep{Yun2001} for typical spectral index 0.7.  Most local galaxies
have SFR much lower than this.  SFR = 200~\Msun~yr$^{-1}$ would
correspond to $L(FIR)>10^{12}$~\Lsun\ \citep{Kennicutt1998}, i.e., to
a ULIRG, which galaxies are very rare locally.  However, the ULIRG
abundance is much higher at high redshift (e.g.,
\citealt{LeFloch2005} Fig.~14; \citealt{Magnelli2011} Fig.~9b), and
the detection of 8234  
$z\le1.2$ LIRGs in the COSMOS field \citep{Feruglio2010} suggests
that there could be of order 20 ULIRGs at $z<1.2$ in the present
sample. Figure~\ref{f:zp} shows observed radio luminosities versus
redshift for sample galaxies with known redshifts.  At $z\le1.2$,
there are 87 sample galaxies with $L({\rm 3~GHz}) > 2 \times
10^{23}$~W~Hz$^{-1}$. Of these, 14 have $L({\rm 3~GHz})$ an order of
magnitude or more above this limit and are unlikely to be star
forming, but the sheer numbers of such luminous galaxies also
suggests that most are AGNs.  The two objects with highest radio
luminosity (at $z\le1.2$: 017 and 193) are radio doubles, clarifying the
presence of an AGN at least in them.  Over all redshifts, 27 sources have $L({\rm
  3~GHz}) > 5\times 10^{24}$~W~Hz$^{-1}$, above any known value for
star formation \citep{Chapman2010}.  The \citet{Huang2009,Huang2012a}
sources should by selection be among the most luminous star formers,
and their luminosities are all between 0.8 and
$4\times10^{24}$~W~Hz$^{-1}$.  It is therefore likely but not certain
that the radio emission from the most luminous radio sources comes
from an AGN, but radio luminosity alone is an uncertain criterion.

The MIR colors themselves can indicate the nature of the sources
\citep{Stern2005, Donley2011} but not with perfect
reliability. \citeauthor{Donley2011} found almost 40\% of powerful
radio galaxies from the SHzRG sample \citep{Seymour2007} outside
their AGN selection region and fully 2/3 of $z>1$ 3CRR radio galaxies
(as distinguished from 3CRR quasars) outside it.  Apparently some
combination of obscuration of the MIR AGN emission by dust and
veiling by the host galaxy can make the overall colors look like a
normal or star-forming galaxy, especially if the host galaxy is
luminous compared to the AGN.  Worse, purely star-forming galaxy SEDs
(at least as represented by some templates) can enter the
\citet{Stern2005} AGN color-color region at some redshifts and for
some values of dust reddening \citep{Donley2011}.  The
\citeauthor{Donley2011} criteria were designed to avoid this
contamination but at the price of missing some AGNs.  In the present
sample, 97 objects (19\%) have MIR colors in or very near the
\citeauthor{Donley2011} AGN box.\footnote{Specifically, the count
  includes all objects with $[3.6]-[5.8]>0.2$ and
  $[4.5]-[8.0]>0.375$.  Given their redshifts as indicated in
  Fig.~\ref{f:donley}, the objects meeting these criteria are likely
  to be AGNs rather than star-forming contaminants.} \citet{Park2010}
found 31 of these to be power-law galaxies, which are clearly AGNs.
(The \citeauthor{Park2010} radio sample was the same one used in this
paper.) Radio luminosities for the 83 objects with redshifts and AGN
MIR colors range from $3\times10^{20}$ to
$8\times10^{25}$~W~Hz$^{-1}$, again showing that radio luminosity by
itself is a poor criterion for source type.

\begin{figure}
\epsscale{1.0}
\vspace*{-1.5in}
\plotone{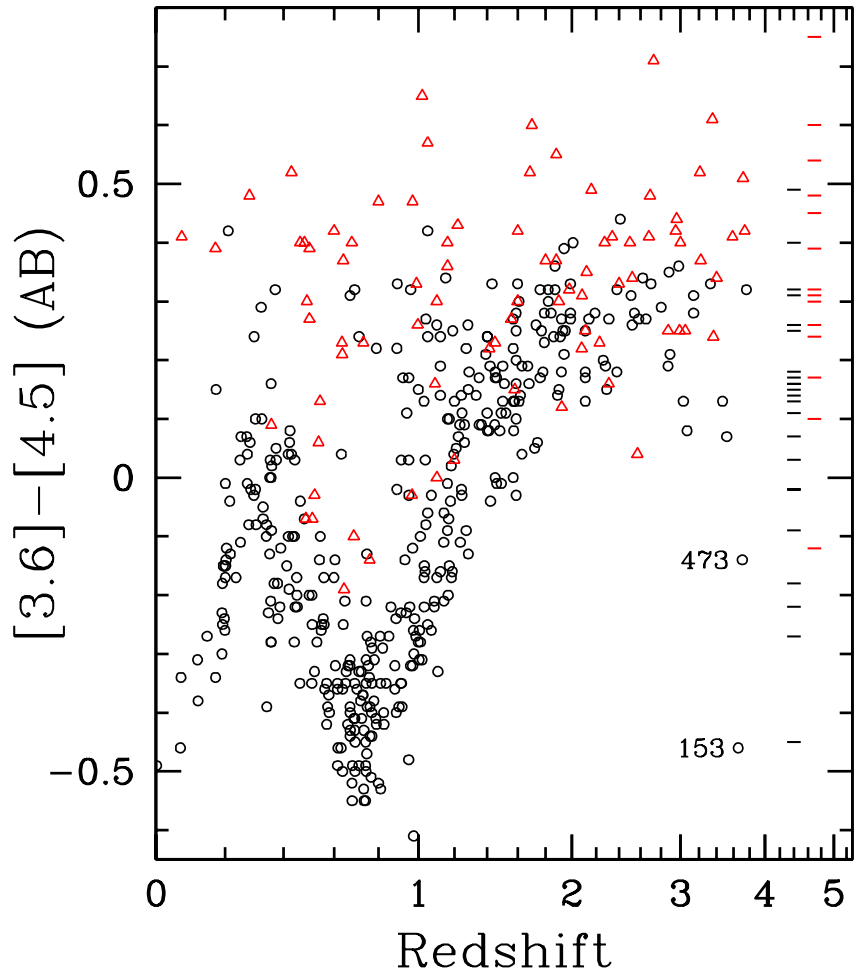}
\caption{IRAC $[3.6]-[4.5]$ color versus redshift (plotted as
  $\log(1+z)$ for clarity). Red triangles denote
  objects with MIR colors indicating AGN emission, and black circles
  denote  other sources.  Colors of sources
  without redshifts are 
  shown as dashes, plotted arbitrarily at redshift 4.7 for those with
  AGN colors  and 4.4 for other objects. Two outliers are labeled
  with their  source nicknames from Table~\ref{t:iracpos}.}
\label{f:z12}
\end{figure}

The galaxies that lack redshifts have MIR colors consistent with AGN
emission or high redshift or both.  In Figure~\ref{f:cc1} in
particular, almost all the galaxies without redshifts are either in
the AGN wedge or just to its left where the $z>1.5$ templates (and
most of the 3CR radio galaxies) lie. This is confirmed more directly
by Figure~\ref{f:z12}, which shows that the galaxies without
redshifts almost all have $[3.6]-[4.5] >0.1$.  This color is
consistent with $z\ga1.1$ or AGN emission or both. The objects
without redshifts tend to be the faintest ones in the sample, as
shown in Figure~\ref{f:rhist}.  This will make it difficult to obtain
redshifts but is consistent with their being obscured AGNs or at
$z>>1$. \citet{Higdon2005} studied a sample of``optically invisible
radio sources'' (OIRSs) with $R\ga26.0$ and $I\ga25.6$. The surface
density of such sources \citep{Higdon2005} predicts more than 25 such
objects in the present sample (which is slightly deeper in
$S(1.4~{\rm GHz})$ than the \citeauthor{Higdon2005} sample), and in
fact there are 46 objects with $R>26$. \citeauthor{Higdon2005}
suggested that the vast majority (87\%) of such sources are AGNs based
on the lack of 24~\micron\ detections, and \citet{Houck2005} gave
evidence that even some OIRSs with 24~\micron\ detections have
spectra consistent with being powered by an AGN.  Half of the OIRSs
in the present sample have MIR colors outside  the AGN boxes in
Figures~\ref{f:donley} and~\ref{f:cc1}, and it seems likely that many
of these are AGN-powered despite their MIR colors.

Even when the source of the radio emission is an AGN, the AGN
emission need not dominate the rest-frame near infrared emission.
Figure~\ref{f:z12} shows that for the majority of the sample,
$[3.6]-[4.5]$ is consistent with a stellar population at the observed
redshift.  Figures~\ref{f:donley} and~\ref{f:cc1} show many radio
source counterparts inside the AGN regions but most outside.  Thus a
substantial minority of radio source counterparts show AGN emission
in the infrared, but most do not.  The origin of the radio emission
from these sources is unclear, and detailed followup will be needed
to determine which galaxies are star forming and which are AGNs.

Authors studying other \mmjy\ radio samples have disagreed on the origin of
the radio emission, although it has always been clear that AGNs and
star forming galaxies are both present
\citep{Benn1993}. \citet{Barger2007} used visible spectroscopy and
X-ray emission to separate star formers from AGNs and concluded that
the majority of the $z<1$ sample are star formers while at least 1/3
of the $L({\rm 3~GHz}) > 2.6 \times 10^{23}$~W~Hz$^{-1}$ sample
(mostly at $z>1$) have X-ray luminosity characteristic of AGNs.  A
substantial allowance for X-ray obscured AGNs
\citep[e.g.,][]{Donley2005} must be added to the numbers of that last
group. \citet{Chapman2010} detected far-infrared emission
characteristic of star formation in at least 40 galaxies of a
sub-sample of 68 radio sources above the same radio luminosity limit,
but the requirement to have a spectroscopic redshift may have biased
their sample against AGNs, and \citet{Moric2010} give reason for
caution in interpreting FIR fluxes.  \citet{Padovani2011} classified
63\% of their radio source counterparts as AGNs based on a complex
scheme involving a wide variety of data.  They also found that the
AGN fraction depends strongly on the radio flux density limit and on
redshift.

All in all, about 10--15\% of the AEGIS20 sample consists of nearby
star-forming galaxies, and about 20--25\% are AGNs, mostly at $z>1$.
The rest are a mix of types, but the exact proportions are still unclear.
Because the sample has essentially complete
counterpart identification, it should be valuable in determining the
origin of the radio emission and the fraction of obscured AGNs at
high redshift.  For comparison with other samples, the median
redshift of the AEGIS20 sample is 1.03 (assuming the unmeasured
redshifts are above this value). This is  lower than that of
the slightly deeper ($S_{1.4}>43$~\mmjy)  \citet{Mainieri2008}
sample, which has a median redshift of 1.18 
with 92\% redshift completeness \citep{Padovani2011}.  That degree of
redshift completeness is vital; an earlier study \citep{Padovani2009}
of the \citeauthor{Mainieri2008} sample with only 70\% redshift
completeness found a median redshift of 0.67.

\subsection{The $K$--$z$ Relation and Stellar Mass}

\begin{figure}
\epsscale{1.0}
\plotone{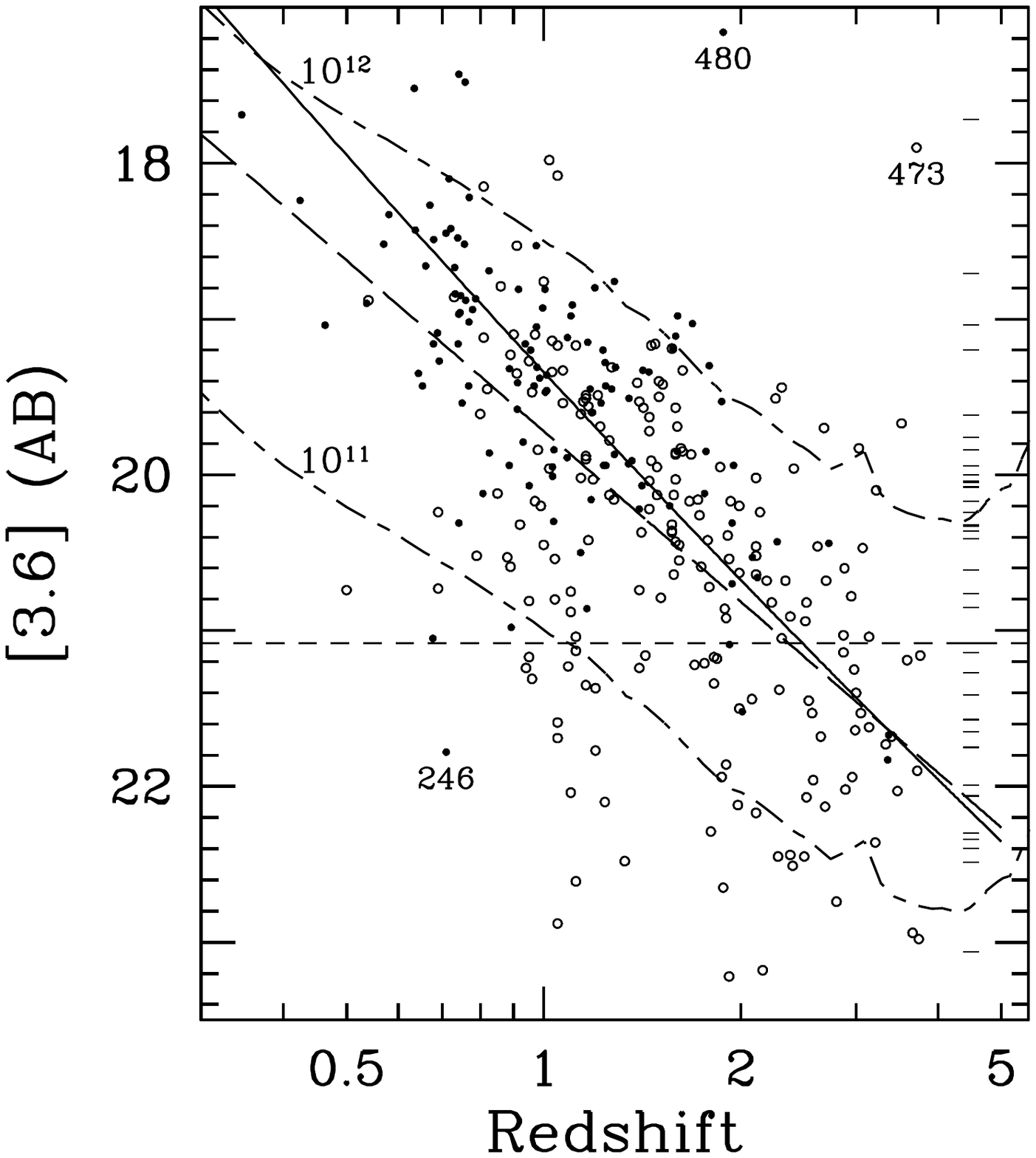}
\vspace*{-1.5in}
\vspace*{-0.75in}
\caption{IRAC 3.6\,\micron\ magnitude as a function of
  redshift. Filled circles denote objects with spectroscopic
  redshifts, and open circles denote objects with photometric
  redshifts. Objects with $L({\rm 3~GHz})<10^{23}$~W~Hz$^{-1}$ are
  omitted; they are likely to be local starburst galaxies. Three
  outlying points are labeled with their source nicknames from
  Table~\ref{t:iracpos}.  Magnitudes of objects with no redshift
  measurement are shown as dashes plotted at an arbitrary redshift of
  4.5.  The horizontal dashed line shows the 90\% completeness limit
  of the IRAC catalog \citep{Barmby2008}.  The solid curve shows the
  flux density from the $K$--$z$ relation given by
  \citet{Willott2003}, i.e., assuming constant flux density from
  observed 2.2 to 3.6~\micron.  The long-dashed line is the same but
  using the \citet{Bryant2009} relation. The two short-long dashed
  lines show the flux density of instantaneous bursts of star
  formation occurring at $z=6$ with initial masses $10^{11}$ and
  $10^{12}$~\Msun, respectively. (The actual stellar mass at relevant
  redshifts is 25--30\% less because of assumed mass loss.)  The
  models come from \citet{bc03}, use a Salpeter IMF, and include no
  dust extinction.}
\label{f:z1}
\end{figure}

Several studies \citep[e.g.,][]{Willott2003,Rocca2004,Bryant2009}
have noted a strong correlation between redshift and the observed
2.2~\micron\ flux density.  However, these studies have all involved
samples with limiting radio flux densities $\ga$200~mJy.
Figure~\ref{f:z1} shows the 3.6~\micron\ magnitude versus redshift
for the current sample, for which few 2.2~\micron\ ($K$) magnitudes
are available.  The wavelength difference between $K$ and [3.6] will
introduce a modest, redshift-dependent offset in the relation, but
the basic form should not change because the dominant radiation
source in both wavelengths is stars in the host galaxy.  Indeed
Figure~\ref{f:z1} shows that the [3.6]--$z$ data are distributed
around the known $K$--$z$ relation but with large scatter.  The
scatter is larger for sources with photometric redshifts but still
substantial even for objects with spectroscopic redshifts. Using a
synthetic $K$ magnituce $K_{synth} \equiv [3.6] + a([3.6]-[4.5])$,
where $a$ is a constant of order unity, does not dramatically reduce
the scatter. The two objects (233, 480) with [3.6] far brighter than
the standard relation are likely to be QSOs, which are normally
excluded from the $K$--$z$ relation. The scatter is far too large to
use the 3.6~\micron\ magnitude as a way to determine redshift for
individual objects, but as a group, the objects without measured
redshifts have magnitudes consistent with $z>1.1$.

The observed 3.6~\micron\ magnitudes of the radio source counterparts
give an indication of the stellar masses.  At the redshifts of
interest, radiation observed at 3.6~\micron\ was emitted near the
stellar radiation peak at 1.6~\micron, and this is closely associated
with the mass of the stellar population
\citep[e.g.,][]{Bell2001}. Figure~\ref{f:z1} shows the magnitudes
that would be observed for single stellar populations that form
instantaneously at $z=6$ and evolve passively thereafter
\citep{bc03}.  Most of the radio source counterparts fall between the
lines for $10^{11}$ and $10^{12}$~\Msun\ but with substantial numbers
below the $10^{11}$~\Msun\ line.  If the populations formed at $z<6$,
the actual galaxy masses could be up to a factor of 3--5 smaller than
indicated.  Using Chabrier instead of Salpeter IMF would lower the
masses by a factor of 1.7 \citep{Bundy2006}.  Furthermore, some of
the 3.6~\micron\ light could come from an AGN rather than a stellar
population; this is almost certainly the case for the extreme
galaxies such as 480 and 473.  On the other hand, the stellar
population models do not include dust extinction, which could cause
the masses to be underestimated by an unknown amount.  Model-fitting
could give better mass estimates for individual galaxies, but the
overall picture is that the radio sources live mostly in massive
galaxies, but the range of stellar masses represented in the sample
is larger than an order of magnitude.

\section{Conclusions}
\label{s:conc}

IRAC images are a powerful means of identifying and classifying radio
sources.  Images with $\sim$9~ks IRAC depth giving
$\sigma\approx0.1$~\mmjy\ \citep{Barmby2008} are sufficient to detect
essentially all counterparts of radio sources in a sample with
1.4~GHz brightness limit of 50~\mmjy/beam.  Radio sources at this
depth are roughly 10--15\% local ($z\la0.5$) starbursts and 20--25\% AGNs mostly
at $z>1$ with the remainder of uncertain nature.  More than 1/3 of the
sample have counterparts with $R_{\rm AB}>24$, and 15\% have $R_{\rm
  AB}>25.5$.  These sources would be very difficult to identify in
$R$-band surveys, and if simple position matching were used, many
would be incorrectly identified with brighter objects that are nearby
on the sky but unrelated to the radio source.

The AEGIS20 sample, now essentially 100\% identified, offers a great
opportunity for detailed studies of the radio population.  X-ray
observations and additional spectra should enable a better
classification between star-forming and active-nuclei galaxies, and
the sample should yield luminosity functions and evolutionary history
of the various populations.




\acknowledgments

The authors thank Jennifer Donley for valuable discussions on
interpretation of IRAC color-color plots and Kate Whittaker for
drawing our attention to and help with the NMBS survey.  We thank
Satoshi Miyazaki for providing the Subaru $R$ images.  We also thank
Pablo P\'erez Gonz\'alez for help using the Rainbow Navigator query
system and especially for his and Guillermo Barro's work to create
the Rainbow database.
This work is based in part on observations made with the Spitzer
Space Telescope, which is operated by the Jet Propulsion Laboratory,
California Institute of Technology under a contract with
NASA. Support for this work was provided by NASA through an award
issued by JPL/Caltech.  The National Radio Astronomy Observatory is a
facility of the National Science Foundation operated under
cooperative agreement by Associated Universities, Inc.
A.L.C. acknowledges  funding from NSF CAREER grant AST-1055081.  DEEP
spectroscopy is supported by the National Science Foundation grants
AST-0071198, AST-0507483, and AST-0808133. This study makes use of
data from AEGIS, a multiwavelength sky survey conducted with the
Chandra, GALEX, Hubble, Keck, CFHT, MMT, Subaru, Palomar, Spitzer,
VLA, and other telescopes and supported in part by the NSF, NASA, and
the STFC. This work has made use of the Rainbow Cosmological Surveys
Database, which is operated by the Universidad Complutense de Madrid
(UCM).  IRAF is distributed by the National Optical Astronomy
Observatory, which is operated by the Association of Universities for
Research in Astronomy (AURA) under cooperative agreement with the
National Science Foundation.



Facilities:
\facility{Spitzer/IRAC}
\facility{Spitzer/MIPS}
\facility{Keck}
\facility{Subaru}
\facility{VLA}



\clearpage
\clearpage

\vspace*{-1in}
\includegraphics[width=\textwidth]{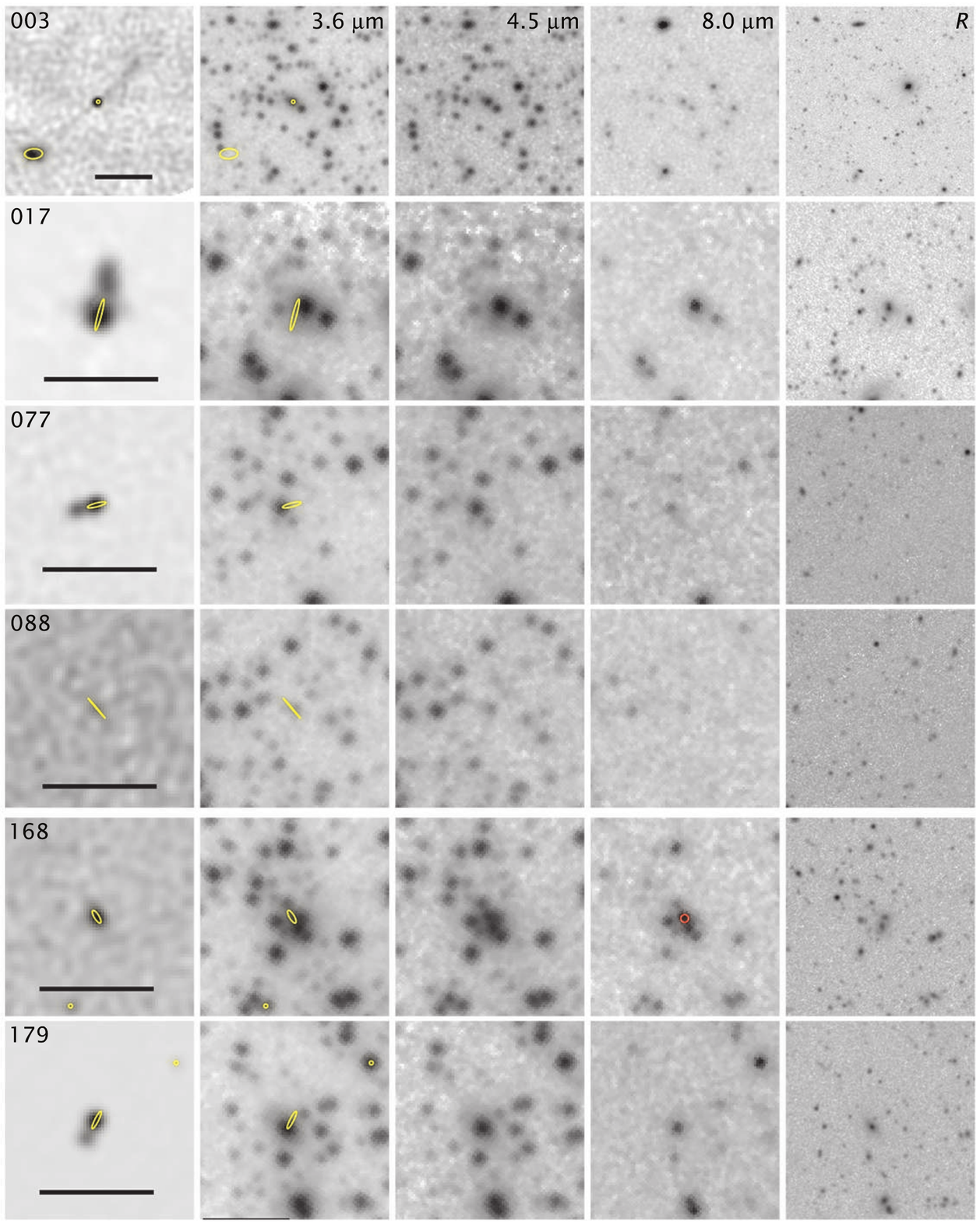}\protect\eject
\vspace*{-1in}
\includegraphics[width=\textwidth]{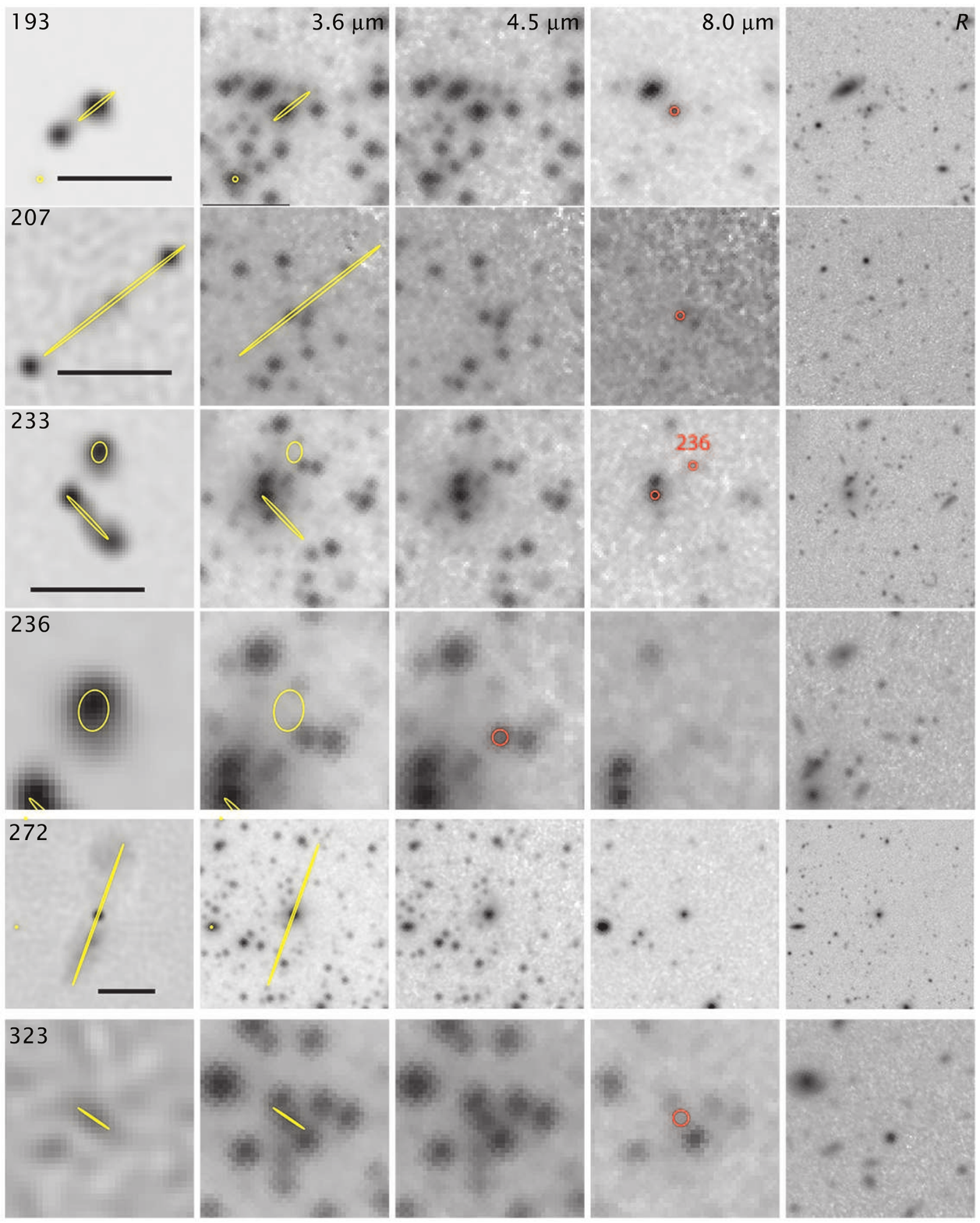}\protect\eject

\clearpage
\vspace*{-1in}
\includegraphics[width=\textwidth]{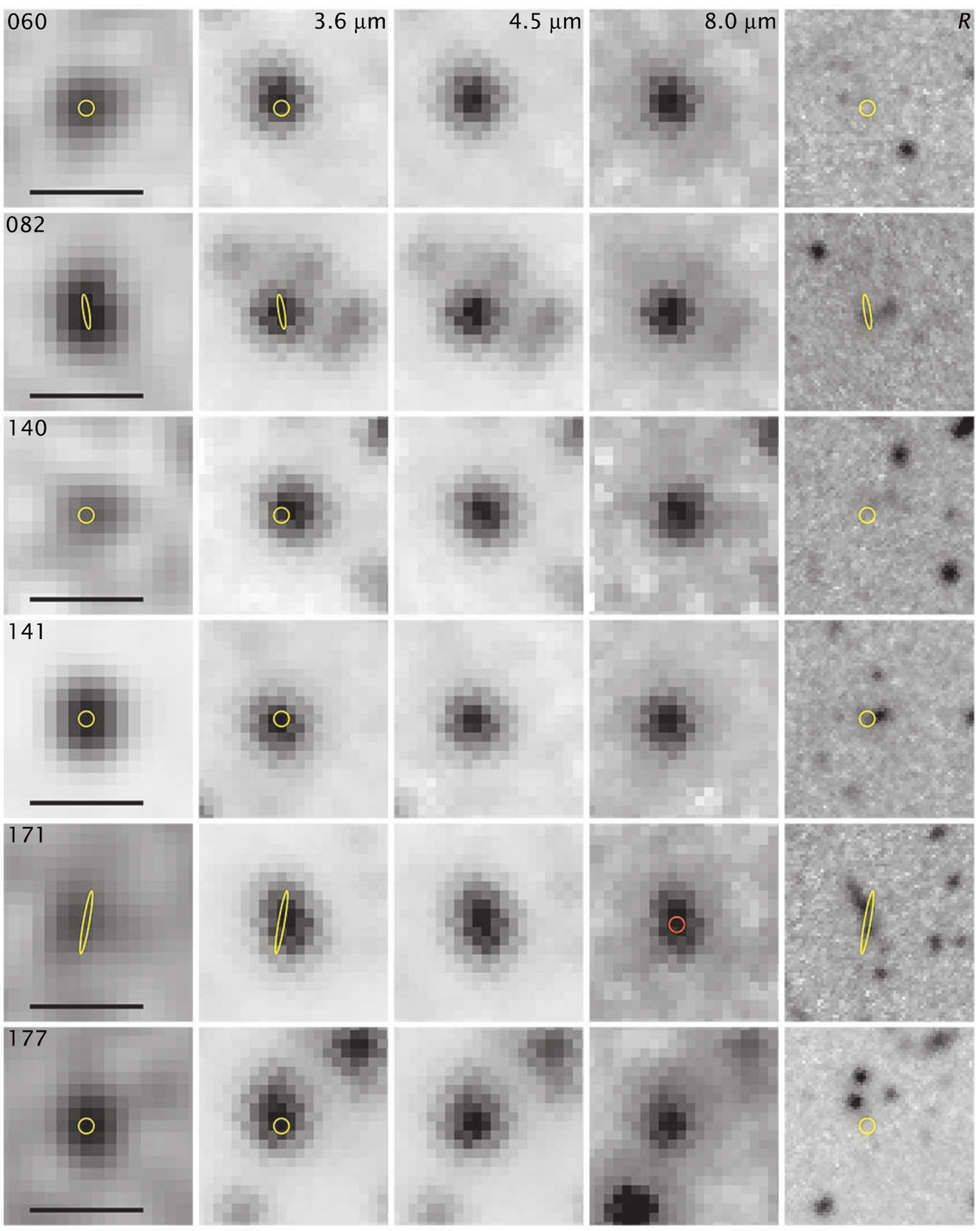}\protect\eject

\clearpage

\textheight=9.0in
. Make sure there is at least one \tablenotemark

\tablenotetext{a}{Tabulated uncertainties in flux density are
  1$\sigma$. Values of $-1$ in integrated uncertainty  indicate
  that the source was poorly fit by a Gaussian.  In these cases,
  integrated flux density was measured in a square that included the
  entire source, and minor axis sizes are not given.}
\tablenotetext{b}{Value $-1$ indicates source unresolved in the minor
  axis direction.}


\end{deluxetable}

\begin{table}
\begin{center}
\caption{\protect\centering
  Radio Sources with No IRAC Identification\label{t:noid}}
\medskip
. Make sure there is at least one \tablenotemark

\vspace*{-4pt}
\tablenotetext{a}{Radial offset from IRAC to radio position.  Where a
  source name is shown, IRAC names and
  positions are from the published catalog 
  \citep{Barmby2008}.  For counterparts not in the published catalog,
  source positions were measured on the published images.}
\tablenotetext{b}{Number of useful images in whichever IRAC
  wavelength has the fewest in
  units of 200~s (one image at 3.6--5.8~\micron; 4 images 
  at 8.0~\micron; see \citealt{Barmby2008}.}
\tablenotetext{c}{Source name in Rainbow database \citep{Barro2011a}}
\tablenotetext{d}{Radial offset from IRAC position measured here to
  source position in Rainbow database.}
\end{deluxetable}
\clearpage
}

\begin{deluxetable}{cccccccccc}
\tabletypesize{\footnotesize}
\tablecaption{$R$ Magnitudes and Redshifts of Radio Source Counterparts\label{t:rmags}}
\tablewidth{0pt}
\tablehead{
\colhead{ID} &  
\colhead{RA\tablenotemark{a}} &
\colhead{Dec\tablenotemark{a}}&
\colhead{Offset\tablenotemark{b}} &
\colhead{Mag type\tablenotemark{c}} &
\colhead{$R$} &
\colhead{$\pm$\tablenotemark{d}} &
\colhead{$z$} &
\colhead{$z$ source\tablenotemark{e}} &
\colhead{$L(3~{\rm GHz})$}
}
\startdata
001&\no&\no&\no&5&23.77&0.02&0.95&6&3.6E+23 \\
003&214.38594&52.53426&0.23&1&23.13&0.01&0.9859&2&2.2E+24 \\
004&214.54142&52.55020&0.28&1&25.14&0.09&2.87&6&7.9E+25 \\
005&214.47884&52.55183&0.12&1&25.30&0.10&1.84&6&1.6E+24 \\
006&214.59998&52.55628&0.43&1&23.35&0.02&1.20&6&6.5E+23 \\
007&214.42169&52.56540&0.28&1&23.73&0.02&3.3630&5&4.1E+25 \\
008&214.56735&52.58665&0.18&4&20.95&0.01&0.3264&2&2.1E+22 \\
009&214.47244&52.59435&0.12&3&21.27&0.01&0.6791&2&1.7E+24 \\
\enddata
\tablenotetext{a}{J2000 position of counterpart if measured
  independently on $R$ image.}
\tablenotetext{b}{Distance in arcseconds from $R$ to IRAC
  position of counterpart.}
\tablenotetext{c}{Magnitude type: 1--3 = SExtractor aperture magnitudes in
  1\farcs53, 2\farcs14, or 3\farcs06 radii respectively, 
  4 =  magnitude in Kron aperture as measured by SExtractor,
  5--7 = {\sc apphot} magnitude  with
  same aperture sizes as 1--3 respectively measured at $R$ position if
  available, otherwise IRAC position.}
\tablenotetext{d}{Magnitude uncertainties given here are statistical
  only, estimated from sky fluctuations.  See text for discussion of
  systematic uncertainties.  ``$-5$'' indicates magnitude given is 5$\sigma$ upper limit.}
\tablenotetext{e}{Redshift source (in priority order if more than one
  redshifts exists):
  1 = DEEP3 (\citealt{Cooper2011,Cooper2012}),
  2 = DEEP2 (\citealt{Davis2003,Newman2012}, http://deep.berkeley.edu/DR3/), 
  3 = MMT  Hectospec \citep{Coil2009}, 
  4 = {\it Spitzer\/}/IRS \citep{Huang2009,Huang2012a}, 
  5 = spectroscopic redshift from Rainbow \citep{Barro2011a},
  6 = photometric redshift from NMBS \citep{Whitaker2011},
  7 = photometric redshift from CFHTLS (\citealt{Coupon2009}:
  http://terapix.iap.fr/cplt/CFHTLS\_deep\_ugriz\_T0004\_zphot\_1.4.tgz),
  8 = photometric redshift from \citet{Huang2012b},
  9 = photometric redshift from Rainbow \citep{Barro2011a}.}

\end{deluxetable}

\begin{deluxetable}{cccccccccc}
\tablecaption{IRAC Magnitudes \label{t:iracmags}}
\tablewidth{0pt}
\tablehead{
\colhead{ID} & 
\colhead{Mag type\tablenotemark{a}} & 
\colhead{3.6\,\micron} &
\colhead{$\pm$} &
\colhead{4.5\,\micron} &
\colhead{$\pm$} &
\colhead{5.8\,\micron} &
\colhead{$\pm$} &
\colhead{8.0\,\micron} &
\colhead{$\pm$} 
}
\startdata
001&1&21.17&0.01&21.14&0.01&21.34&0.11&21.46&0.12 \\
003&1&19.38&0.00&19.65&0.00&19.85&0.02&20.09&0.03 \\
004&1&21.03&0.01&20.78&0.01&20.56&0.05&20.31&0.04 \\
005&1&21.18&0.01&20.90&0.01&20.87&0.06&21.11&0.08 \\
006&1&21.77&0.02&21.74&0.03&21.57&0.13&21.32&0.11 \\
007&1&21.67&0.02&21.43&0.02&21.24&0.10&20.69&0.06 \\
008&4&18.96&0.01&19.01&0.01&19.34&0.06&18.35&0.03 \\
009&3&18.49&0.00&19.04&0.00&19.16&0.02&19.83&0.04 \\
\enddata

\tablecomments{Data in AB magnitudes.  Uncertainty is statistical
  only and is based on fluctuations in the nearby sky.  It is shown
  as 0.00 when less than 0.005 mag, but see text for discussion of
  systematic errors.}

\tablenotetext{a}{Magnitude type: 1--3 =  aperture magnitudes in
  1\farcs53, 2\farcs14, or 3\farcs06 radii respectively, 
  4 =  magnitude in Kron aperture as measured by SExtractor}

\end{deluxetable}

\end{document}